\documentclass[prl,nofootinbib,reprint,a4paper,twocolumn,showpacs,superscriptaddress,showkeys]{revtex4-1}
  \usepackage{amsmath,amssymb,amsfonts}
  \usepackage[dvips]{graphicx}
  \usepackage{graphicx}
  \usepackage{txfonts}
  \usepackage[T1]{fontenc}
  \usepackage{textcomp}
  \usepackage{graphicx}
  \usepackage{epsfig}
  \usepackage{epstopdf}
  \usepackage{upgreek}
  \usepackage{braket}
  \usepackage{bbold}
  \usepackage{psfrag}
  \usepackage[mathscr]{euscript}
  \usepackage[unicode]{hyperref}
  \hypersetup{
      unicode=true,                                           
      a4paper=true,
      plainpages=false,
      pdffitwindow=true,                                      
      pdftitle={Soliton Interferometry},                      
      pdfauthor={John Lloyd Helm},                            
      pdfsubject={},                                          
      pdfkeywords={Bose--Einstein, condensate, BEC, solitons},
      colorlinks=true,                                        
      linkcolor=blue,                                         
      citecolor=blue,                                         
      filecolor=blue,                                         
      urlcolor=blue                                           
  }
  \newcommand{\eqnrefp}[1]{{[Eq.~(\ref{#1})]}}
  
  \newcommand{\eqnreft}[1]{{Eq.~(\ref{#1})}}

  \newcommand{\figreft}[2]{Fig.~\ref{#1}#2}
  \newcommand{\figsreft}[4]{Figs.~\ref{#1}#2 and \ref{#3}#4}

  \newcommand{\figrefp}[2]{[Fig.~\ref{#1}#2]}

  \newcommand{\imgext}{png}

  \newcommand{\etal}{\emph{et al.}~}
  
  \newcommand{\eye}{i}
  \newcommand{\pa}{\partial}
  \newcommand{\xp}{\,\mathrm{e}^}
  \renewcommand{\d}{{\mathrm{d}}}
  \newcommand{\x}{\boldsymbol{\mathrm{r}}}

  \renewcommand{\bra}[1]{\langle#1|}
  \renewcommand{\ket}[1]{|#1\rangle}
  \renewcommand{\sin}{\mathrm{sin}}
  \renewcommand{\cos}{\mathrm{cos}}
  
  \newcommand{\mub}{\mu_{\mathrm B}}
  \newcommand{\gf}{g_F}
  
  \newcommand{\omP}{\omega_\perp}
  \newcommand{\BIP}{{\bf B}_{{\mathrm I\mathrm P}}}
  
  \newcommand{\BPsi}{{\bf \Psi}}
  \newcommand{\BP}{{B}_{\mathrm q}}
  \newcommand{\BZ}{{B}_{z}}
  
  \newcommand{\BPH}{{\tilde{B}}_{\mathrm q}}
  \newcommand{\BZH}{{\tilde{B}}_{z}}
  
  \newcommand{\BG}{b^\prime}
  
  \newcommand{\dS}{\delta_{S}}
  \newcommand{\Fv}{\mathbf{\bar{F}}}
  \newcommand{\Fva}{{\bar{F}_\alpha}}
\begin{document}
%

  \title{Spin-orbit coupled interferometry with ring--trapped Bose--Einstein condensates.}
  \author{J. L. Helm}
  \affiliation{The Dodd--Walls Centre for Photonic and Quantum Technologies, Department of Physics, University of Otago, Dunedin, New Zealand} 
  \author{T. P. Billam}
  \affiliation{Joint Quantum Center (JQC) Durham--Newcastle, School of Mathematics, Statistics and Physics, 
  \\
  Newcastle University, Newcastle upon Tyne NE1 7RU, United Kingdom}
  \author{A. Rakonjac}
  \affiliation{Joint Quantum Center (JQC) Durham--Newcastle, Department of Physics, Durham University, Durham DH1 3LE, United Kingdom}
  \author{S. L. Cornish}
  \affiliation{Joint Quantum Center (JQC) Durham--Newcastle, Department of Physics, Durham University, Durham DH1 3LE, United Kingdom}
  \author{S. A. Gardiner}
  \affiliation{Joint Quantum Center (JQC) Durham--Newcastle, Department of Physics, Durham University, Durham DH1 3LE, United Kingdom}
  \date{\today}
  
  \begin{abstract}
We propose a method of atom-interferometry using a spinor Bose--Einstein
condensate (BEC) with a time-varying magnetic field acting as a coherent
beam-splitter. Our protocol creates long-lived superpositional counterflow
states, which are of fundamental interest and can be made sensitive to both the
Sagnac effect and magnetic fields on the sub-$\mu$G scale. We split a
ring-trapped condensate, initially in the $m_f=0$ hyperfine state, into
superpositions of internal $m_f= \pm 1$ states and condensate superflow, which are
spin-orbit coupled. After interrogation, relative phase accumulation can be
inferred from a population transfer to the $m_f=\pm 1$ states. The counterflow
generation protocol is adiabatically deterministic and does not rely on
coupling to additional optical fields or mechanical stirring techniques. Our
protocol can maximise the classical Fisher information for any rotation,
magnetic field, or interrogation time, and so has the maximum sensitivity
available to uncorrelated particles. Precision can increase with the
interrogation time, and so is limited only by the lifetime of the condensate.
  \end{abstract}
  
  \maketitle
%

\noindent The endeavour to optimally apply matter-wave interferometry has
generated many proposals and prototypes for ultra-sensitive
rotational~\cite{lenef_1997_prl, gustavson_1997_prl, halkyard_pra_2010,
helm_prl_2015, moxley_pra_2015, nolan_pra_2016}, gravitational or
inertial~\cite{snadden_prl_1998, Peters_met_2001, mcguirk_pra_2002,
chu_nat_1999, muller_prl_2008, muller_nat_2010, altin_njp_2013,
canuel_prl_2006}, and gravity wave~\cite{tino_cqg_2007, dimopoulos_prd_2008,
graham_arxiv_2016} detection protocols. In parallel, optical confinement
potentials allow simultaneous trapping of atoms in different magnetic
sublevels, constituting a spinor condensate~\cite{ho_prl_1998, ohmi_jpsp_1998,
stamper_kurn_rmp_2013, kawaguchi_jpr_2012}.
In addition to their coherent nature, the ability to precisely manipulate
motional and spin degrees of freedom using optical, radio-frequency, and
magnetic fields makes spinor condensates a good candidate for the construction
of an interferometer. We focus on a common path interferometric protocol,
applying it to rotational sensing via Sagnac interferometry (where we note that
our general common-path method is also applicable to zero-area Sagnac
interferometry (ZASI)~\cite{sun_prl_1996, eberle_prl_2010}, an often discussed
alternative to the Michelson geometry for optical gravity-wave
detection~\cite{mizuno_oc_1997, punturoi_cqg_2010,
mavalvala_grg_2010,PhysRevLett.116.061102}).
  
  \begin{figure}[!t]
    \includegraphics[width=\linewidth]{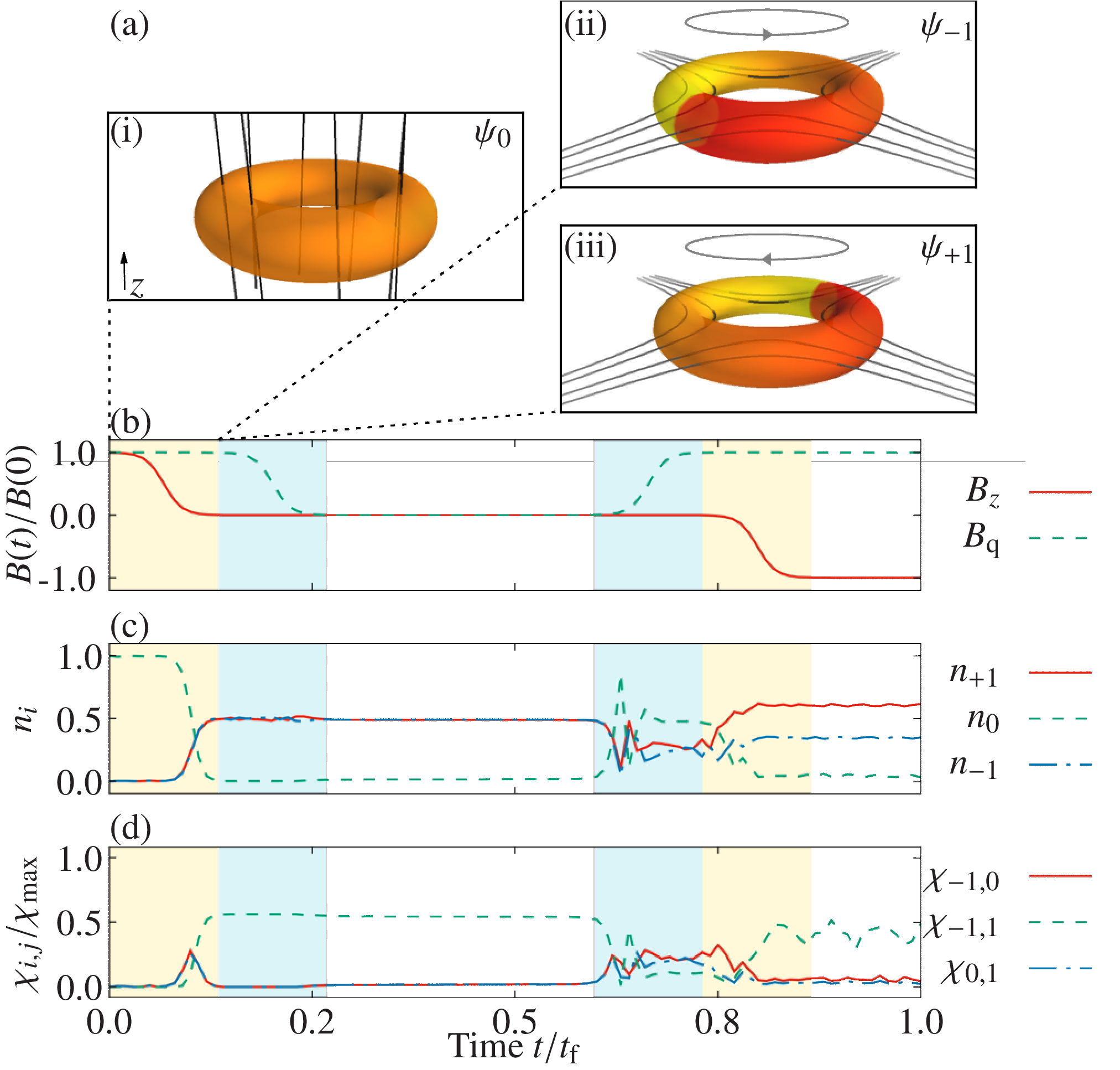}
    \caption{(color online) Overview of spin-orbit coupled interferometry.
    (a) Simulation iso-surface plots at 0.2 of the peak density for the initial condition [$\ket{0}$ in (i)] and immediately after beam-splitting [$\ket{\mp1}$ in (ii) and (iii) respectively]. The color of the iso-surface maps the phase, showing counterflow. Black curves show $B$-field lines. (b) $B$-field ramping scheme. Yellow (outer) shaded regions highlight the beam-splitting processes, while blue (inner) shaded regions show the phase unpinning processes. Numerically calculated norms %
 $n_{i}=\int|\Psi_i|^2\mathop{\d\x}$%
, (c), and overlaps %
 $\chi_{i,j}(t)=\int|\Psi_{i}|^2|\Psi_j|^2\mathop{\d\x}$%
, (d), of the spinor components in the $z$-quantised basis are shown for a $\delta_S=\pi$ interferometry run ending at time $t_{\mathrm f}$.}
    \label{fig:schem}
  \end{figure}
  
In this Letter, we propose a method of matter-wave interferometry in which a
repulsively interacting spinor BEC is split into a superpositional counterflow
state~\cite{PhysRevA.77.053825,PhysRevLett.95.173601} through the use of
topological vortex imprinting~\cite{isoshima_pra_2000, ogawa_pra_2002,
xu_pra_2010, borgh_pra_2016, stamper_kurn_rmp_2013, kawaguchi_jpr_2012}, where
the texture of an externally applied time-varying magnetic field ($B$-field) is
embedded in the condensate's spin and, hence, its phase. In the counterflow
state, each atom is in a superposition of both spin and superflow,
simultaneously moving clockwise and counter-clockwise, while also occupying
multiple hyperfine sub-levels. This class of states is also of fundamental
interest in that it yields superfluid--superfluid counterflow where the
complicating effects of density gradients are substantially
reduced~\cite{PhysRevLett.106.065302}. As the spin and angular-momentum degrees
of freedom are linked, they can be said to be spin-orbit coupled, and we refer
to the method as spin-orbit coupled interferometry (SOCI). This method is
comparable to that proposed by Halkyard \etal~\cite{halkyard_pra_2010} and has
similarly maximised classical Fisher information (denoted $F_{\mathrm C}$)~\citep{haine_prl_2016}.
The procedure uses experimentally accessible time-varying $B$-fields as a
``beam-splitter'' [Fig.~\ref{fig:schem}]. The ``arms'' of the interferometer
are not spatially separate, constituting a common-path interferometer
insensitive to a variety of perturbing factors due to its intrinsic symmetry.
Advantages of our SOCI method are that: (1) our interferometer can
maximize $F_{\mathrm C}$, and so is shown to have the highest sensitivity achievable in
the absence of entanglement; (2) the symmetry and common-path geometry preclude
many systematic phase aberrations; (3) only standard magnetic fields are
required for beam-splitting, with no optical
phase-imprinting~\cite{PhysRevA.77.053825,PhysRevLett.95.173601}, mechanical
stirring or weak
link~\cite{Ragole_prl_2016,PhysRevX.4.031052,PhysRevLett.113.045305} required;
(4) the precision of our interferometer is limited only by the lifetime of the
condensate.

We present our SOCI method analytically in the context of an idealized
measurement, and show that it converts an accumulated phase-difference (due to
rotation at angular frequency $\Omega$, for example) between the counterflowing
components into a difference among the populations, $n_i$, of the spin states.
The measurement sensitivity of such a method can quantified by the classical
Fisher information, $F_{\mathrm C} =\sum_{i}(\pa_\Omega
n_i)^2/n_i$~\citep{haine_prl_2016}.  Using fully-3D numerical simulations of
the spinor mean field, for experimentally realistic parameters, we show that
our method maximises $F_\mathrm{C}$ in the sense that $F_\mathrm{C}$ can be
made equal to the maximum quantum Fisher information, $F_\mathrm{Q}$,
achievable for uncorrelated particles~\citep{haine_prl_2016}.

%

We treat a spin-$F$ condensate as a system of $2F+1$ coupled BECs, working only
in the $z$-quantised (ZQ) representation, where all spin states are labelled
with reference to the $z$-axis. The vector-valued order parameter is
$\BPsi=\sum_{j=-1}^{1}\Psi_j\ket{j}$, where $F_{z}\ket{j}=j\ket{j}$, and $\hbar
F_z/2$ is the $z$-direction angular momentum operator in the ZQ basis.  To
consider the effect of rotations, we introduce an angular momentum
term~\cite{aftalion_pra_2001,mason_pra_2011} characterised by the angular
velocity vector $\bf{\Omega}$; our full mean-field dynamical equations are
then~\cite{stamper_kurn_rmp_2013, kawaguchi_jpr_2012}
  \begin{multline}
    \eye\hbar\frac{\pa}{\pa t}\Psi_j=
    \left[-\frac{\hbar^2}{2m}\nabla^2+V-\eye\hbar({\bf \x}\times{\bf\Omega})\cdot\nabla+g_\mathrm{n}\BPsi^\dagger\BPsi\right]\Psi_j\\
    +\left\{\left[g_\mathrm{s}\Fv \cdot {\bf F} -\mub\gf{\bf B}\cdot{\bf F}\right]\BPsi\right\}_j,\label{eqn:cgpe}
  \end{multline}
where the local spin vector $\Fv$ has components
$\Fva=\Sigma_{j,k}\Psi_k^*\Psi_j\bra{k} F_\alpha\ket{j}$. Here  we have atomic
mass $m$, Bohr magneton $\mub$ and hyperfine gyromagnetic ratio $\gf$ ($=-1/2$
for $^{87}$Rb in the $F=1$ manifold).  The scattering terms are the normal
interaction strength $g_\mathrm{n}$ and spin-spin interaction
strength~$g_{\mathrm{s}}$~\cite{stamper_kurn_rmp_2013}. The
$V=m\omP^2[(\rho-R_0)^2+z^2]/2$ term describes an optical ring
trap~\cite{HRD_NJP_2016}, where $\rho=\sqrt{x^2+y^2}$, giving a radial trapping
frequency $\omP$ and major radius $R_0$. 
While we restrict our analysis to this specfic potential we note that a more general toroidal potential (with density zero at $\rho=0$) could be used to realize a similar interferometer.
Gravity is taken to act in the $z$ direction and does not alter the
symmetry, and so we do not consider it further.  In our numerics we consider
experimental parameters comparable to those described in
\cite{rakonjac_pra_2016,Ray2014}, however, for faster numerics, we take the
radial trapping frequency to be $\omP=2\pi\times80$~Hz, the major radius of the
ring to be $R_0=5a_{\perp}=6.02~\mu$m, and the number of $^{87}$Rb atoms to be
$N=10^4$.
%

  
The idealized behavior of the system can be understood through the eigenvectors of the ${\bf B}\cdot{\bf F}$ operator, in turn determined by the texture of the magnetic field.
Our fundamental requirement is that the $B$-field should have a non-trivial
topology, such that a curve encircling the origin has non-zero winding number,
which is satisfied by either an anti-Helmholtz or Ioffe-Pritchard (IP) coil
configuration. We consider the geometrically simpler IP configuration, which is
quadrupolar in the $x$--$y$ plane.
The Cartesian components can then be written using cylindrical coordinates
$\{\rho,\phi,z\}$ as $\BIP=(\BP(\rho)\cos(\phi),-\BP(\rho)\sin(\phi),\BZ)$,
where the quadrupolar field $\BP(\rho)=\BG\rho$ varies linearly with $\rho$ and
the $z$ bias field is spatially uniform. For $F=1$, in matrix representation
$\ket{1}=(1,0,0)^T$,~$\ket{0}=(0,1,0)^T$, and $\ket{-1}=(0,0,1)^T$:
  \begin{equation}
    {\bf B}\cdot {\bf F}=
    \begin{pmatrix}
      B_z                             &      \BP {\xp{ \eye\phi}}/{\sqrt{2}} & 0                             \\
      \BP {\xp{-\eye\phi}}/{\sqrt{2}} &      0                               & \BP {\xp{\eye\phi}}/{\sqrt{2}}\\
      0                               &      \BP {\xp{-\eye\phi}}/{\sqrt{2}} & -B_z                          
    \end{pmatrix}.\label{eqn:bmat}
  \end{equation}
  The (spatially dependent) eigenvectors of~\eqnreft{eqn:bmat} are:
  \begin{align}
    \ket{\pm {B}}=&([B\pm\BZ]\xp{\eye\phi}, \pm \sqrt{2}\BP, [B\mp\BZ]\xp{-\eye\phi})^T/{2B} ,
  \\
      \ket{Z}=&(-\BP\xp{\eye\phi}, \sqrt{2}\BZ, \BP\xp{-\eye\phi})^T/{\sqrt{2}B}, \label{eqn:zvec}
  \end{align}
  where $B=(\BP^2+\BZ^2)^{1/2}$. The $\ket{+B}$ and $\ket{-B}$ eigenvectors denote the strong- and weak-field-seeking states with eigenvalues $\pm B$, while $\ket{Z}$ is field-insensitive with eigenvalue $0$. Through these eigenvectors we can see the imprinting technique of Ref.~\cite{isoshima_pra_2000}; varying $\BP$ (via $\BG$) and $\BZ$ over time, the condensate remains in a given eigenvector of ${\bf B}\cdot{\bf F}$, but transfers between the $m_f$ states, accumulating $l=F=1$ quantum of angular momentum.
  Some radial dynamics can occur as the $B$-field evolves, but these analytically separate out from the behaviour described by $\mathbf{B}\cdot\mathbf{F}$~\cite{isoshima_pra_2000}, and so are not addressed by our analytics. 
  This implication of spin-gauge symmetry is confirmed by the full 3D numerics.
%

\begin{figure}[!t]
    \includegraphics[width=0.5\textwidth]{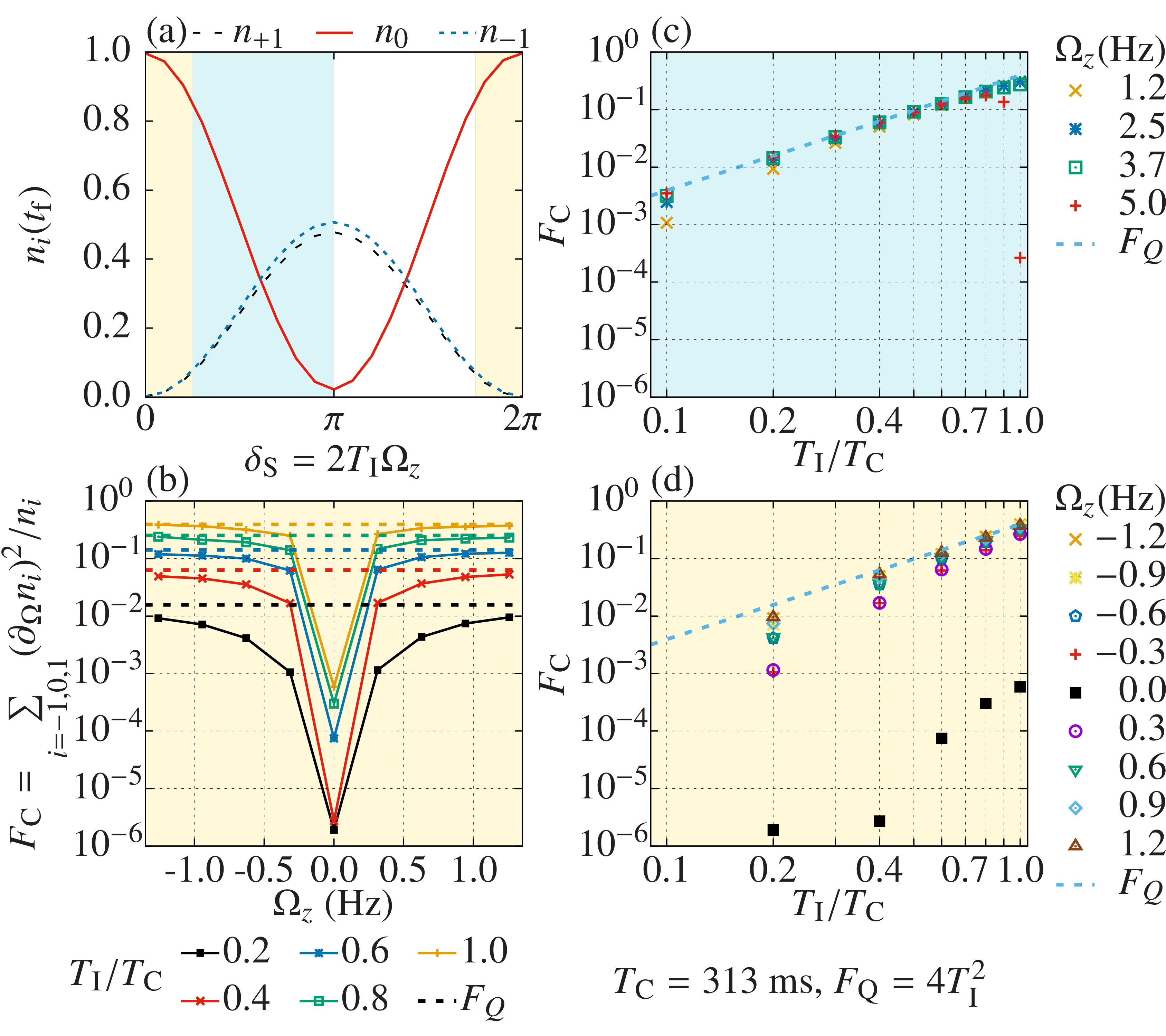}
    \caption{(color online) Results of full 3D numerical simulations quantifying performance of interferometry. (a) Response of the final norm in each component, $n_i(t_f)$, to varying $\Omega_z$. These curves match our analytical result. (b--d) Comparison of quantum ($F_{\mathrm Q}$) and classical ($F_{\mathrm C}$) Fisher information. (b,d) Readings made near the response curve turning-point $\Omega_z=0$ (yellow shading) have suppressed sensitivity ($F_{\mathrm C}<F_{\mathrm Q}$). (c) Readings made on linear segments of the response curve (blue shading) have the maximum sensitivity possible for uncorrelated states ($F_{\mathrm C}=F_{\mathrm Q}$). In all cases we considered $N=10^4$ $^{87}$Rb atoms, with quadrupolar field gradient $b'=37$~Gcm$^{-1}$, initial $z$-bias field $B_z=500$~mG, and field ramping time $T_{s}=3.2$~ms }
    \label{fig:cfi}
  \end{figure}

  To achieve the counterflow state we must first prepare our condensate in the $\ket{0}$ spin-state with a large $z$ bias field $|\BZ|\gg|\BP(R_0)|$. This constitutes the $\ket{Z}$ state. The $\ket{0}$ initial state can be achieved through RF-pumping a $\ket{-1}$ (weak-field-seeking) condensate~\cite{chang_prl_2004}, following transfer to an optical trap, where magnetic trapping is no longer required. 
  With the initial condition fixed in the $\ket{Z}$ state, we obtain the counterflow state by ramping $|\BZ|$ down to zero over a period $T_{\mathrm s}$~[\figreft{fig:schem}], splitting the condensate into a superposition of spin up and spin down [see ~\eqnreft{eqn:zvec}]. We numerically explore two parameter regimes: (1) the quadrupolar field is characterised by an initial gradient $b'=3.7$ G/cm while the initial $z$ bias field is set to $\BZ=50$ mG~[\figreft{fig:schem}] (these parameters are consistent with Ref.~\cite{Ray2014}); (2) we increase the field strengths by a factor of ten to separate the Zeeman and nonlinear timescales, producing a smoother response curve (\figreft{fig:cfi}). We select the ramp-down period $T_\mathrm{s}=32$~ms (or $3.2$~ms for the stronger $B$-field numerics) to be fifty times the Larmor precession time $T_\mathrm{L}=2\pi\hbar/(\mub\gf\BG R_0)=0.64$~ms ($0.064$~ms for the stronger $B$-field numerics), ensuring the spins follow the $B$-field adiabatically. The stronger fields are generally easier to generate experimentally and are easier to vary adiabatically due to their faster associated timescale, putting less stringent requirements on the level of field control. As the atoms in a given spin state have an associated flow field, the condensate is now in a superpositional counterflow state. 
  During counterflow the system is still described by the $\ket{Z}$ eigenstate, and so if we now return $\BZ$ to its initial value (or any other value of suitably large magnitude) the entire condensate will return to $\ket{0}$. We use this method for recombination in our interferometry protocol.
%

  With our beam-splitting and recombination protocols established, we can now consider the impact of a relative phase-shift, consistent with the approach described in \cite{halkyard_pra_2010}. Artificially imprinting a relative phase difference $\delta$ between the spin up and spin down components after the split (at some time when $\BZ=0$), we can re-write our counterflow state as a combination of all three eigenstates of ${\bf B} \cdot {\bf F}$. 
  \begin{align}
    \ket{\Psi}&=\frac{1}{\sqrt{2}}\left(-\xp{\eye\left(\phi+\delta/2\right)}, 0, \xp{-\eye\left(\phi+\delta/2\right)}\right)^T \label{eqn:signal}\\
    &=     \sqrt{\frac{1+\cos(\delta)}{2}}\ket{Z}
      -\eye\sqrt{\frac{1-\cos(\delta)}{2}}\left(\frac{\ket{+B}+\ket{-B}}{\sqrt{2}}\right)\nonumber.
  \end{align}
  Ramping $\BZ$ back up effects the recombination, after which the $\ket{\pm B}$ eigenvectors are the $\ket{\pm 1}$ states, while the $\ket{Z}$ eigenvector is the $\ket{0}$ state. Hence, projecting our final state onto the zero spin state via the $\ket{0}\bra{0}$ projector 
  we obtain an interferometric signal based on the condensate fraction in the $\ket{0}$ state, i.e., $\int|\Psi_0(t_\mathrm{f})|^2\mathop{\d\x}=[1+\cos(\delta)]/2$, where $t_{\mathrm f}$ is the time when our interferometry protocol ends. The populations of the different spin components can be observed experimentally by applying a field gradient in the $z$ direction, resulting in Stern--Gerlach separation.
%

  We now consider prospective interferometry applications, where $\Omega_z$ or $\BZ$ are non-zero during the interrogation counterflow period $T_{\mathrm I}$. We assume the counterflow state is well-described by~\eqnreft{eqn:signal}, discarding structure and dynamics in the $\rho$ and $z$ directions; this assumption is validated by numerical simulation. If we apply the $\eye\hbar(\x\times{\bf \Omega})\cdot\nabla$ operator to each spin component of the counterflow state~\eqnrefp{eqn:signal}, this yields eigenvalues $\pm\hbar\Omega_z$ in the $\ket{\pm1}$ components respectively, and $0$ in the $\ket{0}$ component. These eigenvalues can be incorporated into the diagonal elements of~\eqnreft{eqn:bmat}, combining the rotational and Zeeman terms of~\eqnreft{eqn:cgpe}. The effect of the rotation is simply to offset the strength of the $z$ bias field, which is suppressed (enhanced) as the coordinate system rotates with (against) the magnetic dipole precession. This gauge transformation can be expressed as $\BZH\rightarrow\BZH+\tilde{\Omega}_z$, where ${\bf\tilde{B}}=\mub\gf{\bf B}/\hbar\omP$ and ${\bf \tilde{\Omega}}={\bf \Omega}/\omP$ are dimensionless quantities. 
  Experimentally, $\BZH\gg\BPH\gg\tilde{\Omega}_z$ is typically achievable (and implicit in considering the rotation to be a ``small effect''). 
  Such a transformed system has analogous transformed eigenvectors. Hence, ramping down $|\BZ|\rightarrow0$,we are still in the $\ket{Z}$ eigenstate of $\mathbf{B}\cdot\mathbf{F}$, and therefore expect no accumulation of phase difference between the $\ket{\pm1}$ components.
  In order to observe relative phase accumulation, we must have a superpositional counterflow state in the absence of a quadrupolar field. Carefully ramping down $\BP$ with $\BZ=0$ achieves this aim and unpins the phases of the counterflowing components, but if some small residual contribution $B_{z{\mathrm R}}\neq0$ remains then the system may return to $\ket{0}$ over a slow $\BP$ ramp down as dictated by~\eqnreft{eqn:zvec}, and counterflow is lost. To avoid this restorative effect we must choose the ramp-down curve such that the $\BP$ switch-off is diabatic in some sense. For example, it could be smoothly decaying at first and then cut off instantaneously before the point where $\BP\eqsim10\times B_{z{\mathrm R}}$, or be fully continuous but ramped over a suitably fast timescale.
  The key consideration should be reduction of the radial dynamics and heating associated with diabatic processes, noting that the smaller the residual field, the smaller the associated Zeeman energy, and so the less danger of heating. We also highlight that this restorative effect requires the residual field to satisfy $B_{z{\mathrm R}} \ll \BP(R_0)$ ($\sim 1.57$ mG for our weak $B$-field numerics). This upper bound scales linearly with $R_0$ and $b'$, and can easily be raised.
  In the complete absence of magnetic fields (and, by spin-gauge symmetry, rotations) the hyperfine states become degenerate. In general this leads to undesirable spin-flips, which become more energetically allowable as $B\rightarrow0$. Note that spin-flipping collisions are suppressed in the superpositional counterflow state~\cite{Note1}.
  Another possible source of undesirable spin-flips is stray fields. However, assuming the field can be controlled on the mG scale, such processes have long associated timescales and can be ignored. Finally, we note that quantum and thermal fluctuations may be another source of spontaneous spin-flips, but such analysis is beyond the scope of this Letter. A strategy to avoid these spin-flips would be to purposefully retain a nonzero $B_{z{\mathrm R}}$.

  Once the quadrupolar field is absent, the $\ket{\pm1}$ components can evolve freely, and accumulate a Sagnac phase $\dS$ for $\Omega_z\neq0$, or a Zeeman-energy phase $\delta_{\mathrm Z}$ for $B_{z{\mathrm R}}\neq0$. The phase magnitude can be quantified in terms of either the ring's enclosed area or the interrogation period $T_{\mathrm I}$~\cite{halkyard_pra_2010}. Allowing each component to perform the equivalent of one full circulation around the ring produces a Sagnac phase $\dS=4A\Omega_zm/\hbar$~\cite{halkyard_pra_2010,helm_prl_2015}. The particle velocity around the ring is given by a vortex velocity-field, ${\bf v}=(\hbar/m\rho)\hat{\bf \phi}$, and so the time for a single particle to fully circumnavigate the ring (such that $\rho=R_0$) to be $T_{\mathrm C}=2\pi R_0^2m/\hbar=2Am/\hbar$~($=313$~ms for our parameters). The phase accumulated for an arbitrary interrogation time $T_{\mathrm I}$ is then $\dS=(T_{\mathrm I}/T_{\mathrm C})4\Omega_z Am/\hbar =2\Omega_z T_{\mathrm I}$. The same arguments apply for the Zeeman-energy phase under the substitution $\Omega_z\rightarrow(\mub/\hbar)\BZ$.
  After interrogation, restoring the quadrupolar $B$-field projects our phase-shifted wave-function onto the eigenstates of $\mathbf{B}\cdot\mathbf{F}$. As phases are accumulated the populations in the $\ket{\pm B}$, $\ket{Z}$ basis differ upon restoration of the quadrupolar field~\eqnrefp{eqn:signal}. This induces some radial oscillations as the $\ket{\pm B}$ eigenstates are respectively strong and weak-field-seeking. These oscillations can be seen in the overlap integrals shown in Fig.~\ref{fig:schem}(d), however they do not affect the recombination as the radial dynamics analytically decouple from the eigenvectors of $\mathbf{B}\cdot\mathbf{F}$. 
%

  We show results of numerical simulations of~\eqnreft{eqn:cgpe} (using CUDA~\cite{cuda_2008}) in~\figsreft{fig:schem}{}{fig:cfi}{}. In ~\figreft{fig:schem}~we performed an interferometry procedure with Sagnac phase $\delta_{\mathrm S}=\pi$, fixing the interrogation times $T_{\mathrm I}=T_{\mathrm C}$ for complete circulation around the ring, and employing weak $B$-fields consistent with~\citet{Ray2014}. These fields require longer timescales, allowing us to better see the dynamics. Subplots (b), (c) and (d) display the full time evolution of the $B$-field, the norms $n_{i}=\int|\Psi_i|^2\mathop{\d\x}$ of each component, and the density-density overlap integrals $\chi_{i,j}(t)=\int|\Psi_{i}|^2|\Psi_j|^2\mathop{\d\x}$ respectively. There is a small difference between $n_{\pm1}$ after recombination, as the weak fields used in these numerics make the Zeeman and nonlinear timescales comparable, compromising the dynamics. The result is still commensurate with our analytical predictions even in this sub-optimal regime. We observe good overlap during the counterflow phase, verifying that radial dynamics do not affect interrogation and that our method is a good example of a common path interferometer. After restoring the quadrupolar field oscillations are evident in the $\chi_{-1,+1}$ overlap integral as a result of the condensate now populating the field-sensitive $\ket{\pm{B}}$ eigenstates.
  In~\figreft{fig:cfi}(a) we show the response curve obtained by varying $\Omega_z$ while again holding constant the interrogation time $T_{\mathrm I}=T_{\mathrm C}$. For these we used stronger $B$-fields ($b'=37$~Gcm$^{-1}$ and $\BZ=500$~mG). The response curve is smooth and in good quantitative agreement with our prediction~\eqnrefp{eqn:signal}. We report that response curves obtained by varying $B_{z{\mathrm R}}$, for field sensing on the sub-$\mu$G scale, are in good quantitative agreement with those obtained by varying $\Omega_z$. 
  In~\figreft{fig:cfi}(b--d) we show calculations of the classical Fisher information $F_{\mathrm C}=\sum_{i=-1}^{1}(\pa_\Omega n_i)^2/n_i$ and show that for $\Omega_z\neq0$ it is approximately equal to the quantum Fisher information $F_{\mathrm Q}=4l^2T_{\mathrm I}^2$, the upper limit achievable for uncorrelated particles and so the upper limit available to mean-field treatments~\citep{haine_prl_2016}. The counterflow quantization is $l=F=1$ in our method. This confirms that our SOCI protocol maximises $F_{\mathrm C}$ for an arbitrary pre-selected read-off time. 
  
  For $\Omega\sim0$~\figrefp{fig:cfi}{~(d,b)}, the value of $F_{\mathrm C}$ is dominated by the small number count in the $\ket{\pm1}$ modes such that even small deviations from zero are highly undesirable, as is the case with all 2-mode interferometers. Our protocol can be designed to avoid this issue through the addition of an extra set of quadrupole bars to the IP coils. Using the secondary bars during recombination allows an arbitrary rotation of the quadrupole field about the $z$-axis, effecting a coordinate transformation equivalent to a phase shift $\delta$~\eqnrefp{eqn:signal}. Such a phase shift could move the response-curve to a more favourable location with $F_{\mathrm C}=F_{\mathrm Q}$. In this way, any rotation could be measured with precision limited only by the lifetime of the condensate. The sensitivity can be further increased by using a higher $F$ manifold, increasing $l$ and increasing precision.
  Note that the equivalency between $\tilde{\Omega}_z$ and $\BZH$ requires that care be taken in experimental measurements. The maximum value of $\Omega_z=2\pi\times5.0$~Hz used in the numerics corresponds to $\BZ=3.56$~$\mu$G. As such, it should be straightforward to make single-shot field measurements on the sub-$\mu$G scale, as large rotations should be absent. Similarly, a spin-echo technique~\cite{PhysRev.80.580} would allow the exclusion of Zeeman phases~\cite{halkyard_pra_2010}.
%

  In conclusion, we present a BEC interferometry protocol which requires only the careful control of standard $B$-fields and an optical ring-trap. Our protocol gives the greatest possible degree of access to measurement information for uncorrelated systems and, through its maximal spatial overlap, is a good candidate for Heisenberg limited interferometry~\cite{gross_jpb_2012,ferris_njp_2010,Ragole_prl_2016}. We have also presented the results of full 3D multicomponent mean-field calculations of the Fisher information which demonstrate the robustness of our approach in the absence of idealizing approximations.

  The data presented in this Letter can be found in Ref.~\cite{data}.

  We thank A. L. Marchant, R. J. Bettles, C. Weiss, and S. A. Haine for useful discussions, the UK EPSRC (grant number EP/K03250X/1) and the Leverhulme Trust (grant number RP2013-K-009).


\begin{thebibliography}{49}%
\makeatletter
\providecommand \@ifxundefined [1]{%
 \@ifx{#1\undefined}
}%
\providecommand \@ifnum [1]{%
 \ifnum #1\expandafter \@firstoftwo
 \else \expandafter \@secondoftwo
 \fi
}%
\providecommand \@ifx [1]{%
 \ifx #1\expandafter \@firstoftwo
 \else \expandafter \@secondoftwo
 \fi
}%
\providecommand \natexlab [1]{#1}%
\providecommand \enquote  [1]{``#1''}%
\providecommand \bibnamefont  [1]{#1}%
\providecommand \bibfnamefont [1]{#1}%
\providecommand \citenamefont [1]{#1}%
\providecommand \href@noop [0]{\@secondoftwo}%
\providecommand \href [0]{\begingroup \@sanitize@url \@href}%
\providecommand \@href[1]{\@@startlink{#1}\@@href}%
\providecommand \@@href[1]{\endgroup#1\@@endlink}%
\providecommand \@sanitize@url [0]{\catcode `\\12\catcode `\$12\catcode
  `\&12\catcode `\#12\catcode `\^12\catcode `\_12\catcode `\%12\relax}%
\providecommand \@@startlink[1]{}%
\providecommand \@@endlink[0]{}%
\providecommand \url  [0]{\begingroup\@sanitize@url \@url }%
\providecommand \@url [1]{\endgroup\@href {#1}{\urlprefix }}%
\providecommand \urlprefix  [0]{URL }%
\providecommand \Eprint [0]{\href }%
\providecommand \doibase [0]{http://dx.doi.org/}%
\providecommand \selectlanguage [0]{\@gobble}%
\providecommand \bibinfo  [0]{\@secondoftwo}%
\providecommand \bibfield  [0]{\@secondoftwo}%
\providecommand \translation [1]{[#1]}%
\providecommand \BibitemOpen [0]{}%
\providecommand \bibitemStop [0]{}%
\providecommand \bibitemNoStop [0]{.\EOS\space}%
\providecommand \EOS [0]{\spacefactor3000\relax}%
\providecommand \BibitemShut  [1]{\csname bibitem#1\endcsname}%
\let\auto@bib@innerbib\@empty
\bibitem [{\citenamefont {Lenef}\ \emph {et~al.}(1997)\citenamefont {Lenef},
  \citenamefont {Hammond}, \citenamefont {Smith}, \citenamefont {Chapman},
  \citenamefont {Rubenstein},\ and\ \citenamefont
  {Pritchard}}]{lenef_1997_prl}%
  \BibitemOpen
  \bibfield  {author} {\bibinfo {author} {\bibfnamefont {A.}~\bibnamefont
  {Lenef}}, \bibinfo {author} {\bibfnamefont {T.~D.}\ \bibnamefont {Hammond}},
  \bibinfo {author} {\bibfnamefont {E.~T.}\ \bibnamefont {Smith}}, \bibinfo
  {author} {\bibfnamefont {M.~S.}\ \bibnamefont {Chapman}}, \bibinfo {author}
  {\bibfnamefont {R.~A.}\ \bibnamefont {Rubenstein}}, \ and\ \bibinfo {author}
  {\bibfnamefont {D.~E.}\ \bibnamefont {Pritchard}},\ }\href {\doibase
  10.1103/PhysRevLett.78.760} {\bibfield  {journal} {\bibinfo  {journal} {Phys.
  Rev. Lett.}\ }\textbf {\bibinfo {volume} {78}},\ \bibinfo {pages} {760}
  (\bibinfo {year} {1997})}\BibitemShut {NoStop}%
\bibitem [{\citenamefont {Gustavson}\ \emph {et~al.}(1997)\citenamefont
  {Gustavson}, \citenamefont {Bouyer},\ and\ \citenamefont
  {Kasevich}}]{gustavson_1997_prl}%
  \BibitemOpen
  \bibfield  {author} {\bibinfo {author} {\bibfnamefont {T.~L.}\ \bibnamefont
  {Gustavson}}, \bibinfo {author} {\bibfnamefont {P.}~\bibnamefont {Bouyer}}, \
  and\ \bibinfo {author} {\bibfnamefont {M.~A.}\ \bibnamefont {Kasevich}},\
  }\href {\doibase 10.1103/PhysRevLett.78.2046} {\bibfield  {journal} {\bibinfo
   {journal} {Phys. Rev. Lett.}\ }\textbf {\bibinfo {volume} {78}},\ \bibinfo
  {pages} {2046} (\bibinfo {year} {1997})}\BibitemShut {NoStop}%
\bibitem [{\citenamefont {Halkyard}\ \emph {et~al.}(2010)\citenamefont
  {Halkyard}, \citenamefont {Jones},\ and\ \citenamefont
  {Gardiner}}]{halkyard_pra_2010}%
  \BibitemOpen
  \bibfield  {author} {\bibinfo {author} {\bibfnamefont {P.~L.}\ \bibnamefont
  {Halkyard}}, \bibinfo {author} {\bibfnamefont {M.~P.~A.}\ \bibnamefont
  {Jones}}, \ and\ \bibinfo {author} {\bibfnamefont {S.~A.}\ \bibnamefont
  {Gardiner}},\ }\href {\doibase 10.1103/PhysRevA.81.061602} {\bibfield
  {journal} {\bibinfo  {journal} {Phys. Rev. A}\ }\textbf {\bibinfo {volume}
  {81}},\ \bibinfo {pages} {061602} (\bibinfo {year} {2010})}\BibitemShut
  {NoStop}%
\bibitem [{\citenamefont {Helm}\ \emph {et~al.}(2015)\citenamefont {Helm},
  \citenamefont {Cornish},\ and\ \citenamefont {Gardiner}}]{helm_prl_2015}%
  \BibitemOpen
  \bibfield  {author} {\bibinfo {author} {\bibfnamefont {J.~L.}\ \bibnamefont
  {Helm}}, \bibinfo {author} {\bibfnamefont {S.~L.}\ \bibnamefont {Cornish}}, \
  and\ \bibinfo {author} {\bibfnamefont {S.~A.}\ \bibnamefont {Gardiner}},\
  }\href {\doibase 10.1103/PhysRevLett.114.134101} {\bibfield  {journal}
  {\bibinfo  {journal} {Phys. Rev. Lett.}\ }\textbf {\bibinfo {volume} {114}},\
  \bibinfo {pages} {134101} (\bibinfo {year} {2015})}\BibitemShut {NoStop}%
\bibitem [{\citenamefont {Moxley}\ \emph {et~al.}(2016)\citenamefont {Moxley},
  \citenamefont {Dowling}, \citenamefont {Dai},\ and\ \citenamefont
  {Byrnes}}]{moxley_pra_2015}%
  \BibitemOpen
  \bibfield  {author} {\bibinfo {author} {\bibfnamefont {F.~I.}\ \bibnamefont
  {Moxley}}, \bibinfo {author} {\bibfnamefont {J.~P.}\ \bibnamefont {Dowling}},
  \bibinfo {author} {\bibfnamefont {W.}~\bibnamefont {Dai}}, \ and\ \bibinfo
  {author} {\bibfnamefont {T.}~\bibnamefont {Byrnes}},\ }\href {\doibase
  10.1103/PhysRevA.93.053603} {\bibfield  {journal} {\bibinfo  {journal} {Phys.
  Rev. A}\ }\textbf {\bibinfo {volume} {93}},\ \bibinfo {pages} {053603}
  (\bibinfo {year} {2016})}\BibitemShut {NoStop}%
\bibitem [{\citenamefont {Nolan}\ \emph {et~al.}(2016)\citenamefont {Nolan},
  \citenamefont {Sabbatini}, \citenamefont {Bromley}, \citenamefont {Davis},\
  and\ \citenamefont {Haine}}]{nolan_pra_2016}%
  \BibitemOpen
  \bibfield  {author} {\bibinfo {author} {\bibfnamefont {S.~P.}\ \bibnamefont
  {Nolan}}, \bibinfo {author} {\bibfnamefont {J.}~\bibnamefont {Sabbatini}},
  \bibinfo {author} {\bibfnamefont {M.~W.~J.}\ \bibnamefont {Bromley}},
  \bibinfo {author} {\bibfnamefont {M.~J.}\ \bibnamefont {Davis}}, \ and\
  \bibinfo {author} {\bibfnamefont {S.~A.}\ \bibnamefont {Haine}},\ }\href
  {\doibase 10.1103/PhysRevA.93.023616} {\bibfield  {journal} {\bibinfo
  {journal} {Phys. Rev. A}\ }\textbf {\bibinfo {volume} {93}},\ \bibinfo
  {pages} {023616} (\bibinfo {year} {2016})}\BibitemShut {NoStop}%
\bibitem [{\citenamefont {Snadden}\ \emph {et~al.}(1998)\citenamefont
  {Snadden}, \citenamefont {McGuirk}, \citenamefont {Bouyer}, \citenamefont
  {Haritos},\ and\ \citenamefont {Kasevich}}]{snadden_prl_1998}%
  \BibitemOpen
  \bibfield  {author} {\bibinfo {author} {\bibfnamefont {M.~J.}\ \bibnamefont
  {Snadden}}, \bibinfo {author} {\bibfnamefont {J.~M.}\ \bibnamefont
  {McGuirk}}, \bibinfo {author} {\bibfnamefont {P.}~\bibnamefont {Bouyer}},
  \bibinfo {author} {\bibfnamefont {K.~G.}\ \bibnamefont {Haritos}}, \ and\
  \bibinfo {author} {\bibfnamefont {M.~A.}\ \bibnamefont {Kasevich}},\ }\href
  {\doibase 10.1103/PhysRevLett.81.971} {\bibfield  {journal} {\bibinfo
  {journal} {Phys. Rev. Lett.}\ }\textbf {\bibinfo {volume} {81}},\ \bibinfo
  {pages} {971} (\bibinfo {year} {1998})}\BibitemShut {NoStop}%
\bibitem [{\citenamefont {Peters}\ \emph {et~al.}(2001)\citenamefont {Peters},
  \citenamefont {Chung},\ and\ \citenamefont {Chu}}]{Peters_met_2001}%
  \BibitemOpen
  \bibfield  {author} {\bibinfo {author} {\bibfnamefont {A.}~\bibnamefont
  {Peters}}, \bibinfo {author} {\bibfnamefont {K.~Y.}\ \bibnamefont {Chung}}, \
  and\ \bibinfo {author} {\bibfnamefont {S.}~\bibnamefont {Chu}},\ }\href
  {\doibase 10.1088/0026-1394/38/1/4} {\bibfield  {journal} {\bibinfo
  {journal} {Metrologia}\ }\textbf {\bibinfo {volume} {38}},\ \bibinfo {pages}
  {25} (\bibinfo {year} {2001})}\BibitemShut {NoStop}%
\bibitem [{\citenamefont {McGuirk}\ \emph {et~al.}(2002)\citenamefont
  {McGuirk}, \citenamefont {Foster}, \citenamefont {Fixler}, \citenamefont
  {Snadden},\ and\ \citenamefont {Kasevich}}]{mcguirk_pra_2002}%
  \BibitemOpen
  \bibfield  {author} {\bibinfo {author} {\bibfnamefont {J.~M.}\ \bibnamefont
  {McGuirk}}, \bibinfo {author} {\bibfnamefont {G.~T.}\ \bibnamefont {Foster}},
  \bibinfo {author} {\bibfnamefont {J.~B.}\ \bibnamefont {Fixler}}, \bibinfo
  {author} {\bibfnamefont {M.~J.}\ \bibnamefont {Snadden}}, \ and\ \bibinfo
  {author} {\bibfnamefont {M.~A.}\ \bibnamefont {Kasevich}},\ }\href {\doibase
  10.1103/PhysRevA.65.033608} {\bibfield  {journal} {\bibinfo  {journal} {Phys.
  Rev. A}\ }\textbf {\bibinfo {volume} {65}},\ \bibinfo {pages} {033608}
  (\bibinfo {year} {2002})}\BibitemShut {NoStop}%
\bibitem [{\citenamefont {Chu}\ \emph {et~al.}(1999)\citenamefont {Chu},
  \citenamefont {Peters},\ and\ \citenamefont {Chung}}]{chu_nat_1999}%
  \BibitemOpen
  \bibfield  {author} {\bibinfo {author} {\bibfnamefont {S.}~\bibnamefont
  {Chu}}, \bibinfo {author} {\bibfnamefont {A.}~\bibnamefont {Peters}}, \ and\
  \bibinfo {author} {\bibfnamefont {K.~Y.}\ \bibnamefont {Chung}},\ }\href
  {\doibase 10.1038/23655} {\bibfield  {journal} {\bibinfo  {journal} {Nature}\
  }\textbf {\bibinfo {volume} {400}},\ \bibinfo {pages} {849} (\bibinfo {year}
  {1999})}\BibitemShut {NoStop}%
\bibitem [{\citenamefont {M\"uller}\ \emph {et~al.}(2008)\citenamefont
  {M\"uller}, \citenamefont {Chiow}, \citenamefont {Herrmann}, \citenamefont
  {Chu},\ and\ \citenamefont {Chung}}]{muller_prl_2008}%
  \BibitemOpen
  \bibfield  {author} {\bibinfo {author} {\bibfnamefont {H.}~\bibnamefont
  {M\"uller}}, \bibinfo {author} {\bibfnamefont {S.-w.}\ \bibnamefont {Chiow}},
  \bibinfo {author} {\bibfnamefont {S.}~\bibnamefont {Herrmann}}, \bibinfo
  {author} {\bibfnamefont {S.}~\bibnamefont {Chu}}, \ and\ \bibinfo {author}
  {\bibfnamefont {K.-Y.}\ \bibnamefont {Chung}},\ }\href {\doibase
  10.1103/PhysRevLett.100.031101} {\bibfield  {journal} {\bibinfo  {journal}
  {Phys. Rev. Lett.}\ }\textbf {\bibinfo {volume} {100}},\ \bibinfo {pages}
  {031101} (\bibinfo {year} {2008})}\BibitemShut {NoStop}%
\bibitem [{\citenamefont {M\"{u}ller}\ \emph {et~al.}(2010)\citenamefont
  {M\"{u}ller}, \citenamefont {Peters},\ and\ \citenamefont
  {Chu}}]{muller_nat_2010}%
  \BibitemOpen
  \bibfield  {author} {\bibinfo {author} {\bibfnamefont {H.}~\bibnamefont
  {M\"{u}ller}}, \bibinfo {author} {\bibfnamefont {A.}~\bibnamefont {Peters}},
  \ and\ \bibinfo {author} {\bibfnamefont {S.}~\bibnamefont {Chu}},\ }\href
  {\doibase 10.1038/nature08776} {\bibfield  {journal} {\bibinfo  {journal}
  {Nature}\ }\textbf {\bibinfo {volume} {463}},\ \bibinfo {pages} {926}
  (\bibinfo {year} {2010})}\BibitemShut {NoStop}%
\bibitem [{\citenamefont {Altin}\ \emph {et~al.}(2013)\citenamefont {Altin},
  \citenamefont {Johnsson}, \citenamefont {Negnevitsky}, \citenamefont
  {Dennis}, \citenamefont {Anderson}, \citenamefont {Debs}, \citenamefont
  {Szigeti}, \citenamefont {Hardman}, \citenamefont {Bennetts}, \citenamefont
  {McDonald}, \citenamefont {Turner}, \citenamefont {Close},\ and\
  \citenamefont {Robins}}]{altin_njp_2013}%
  \BibitemOpen
  \bibfield  {author} {\bibinfo {author} {\bibfnamefont {P.~A.}\ \bibnamefont
  {Altin}}, \bibinfo {author} {\bibfnamefont {M.~T.}\ \bibnamefont {Johnsson}},
  \bibinfo {author} {\bibfnamefont {V.}~\bibnamefont {Negnevitsky}}, \bibinfo
  {author} {\bibfnamefont {G.~R.}\ \bibnamefont {Dennis}}, \bibinfo {author}
  {\bibfnamefont {R.~P.}\ \bibnamefont {Anderson}}, \bibinfo {author}
  {\bibfnamefont {J.~E.}\ \bibnamefont {Debs}}, \bibinfo {author}
  {\bibfnamefont {S.~S.}\ \bibnamefont {Szigeti}}, \bibinfo {author}
  {\bibfnamefont {K.~S.}\ \bibnamefont {Hardman}}, \bibinfo {author}
  {\bibfnamefont {S.}~\bibnamefont {Bennetts}}, \bibinfo {author}
  {\bibfnamefont {G.~D.}\ \bibnamefont {McDonald}}, \bibinfo {author}
  {\bibfnamefont {L.~D.}\ \bibnamefont {Turner}}, \bibinfo {author}
  {\bibfnamefont {J.~D.}\ \bibnamefont {Close}}, \ and\ \bibinfo {author}
  {\bibfnamefont {N.~P.}\ \bibnamefont {Robins}},\ }\href {\doibase
  10.1088/1367-2630/15/2/023009} {\bibfield  {journal} {\bibinfo  {journal}
  {New J. Phys.}\ }\textbf {\bibinfo {volume} {15}},\ \bibinfo {pages} {023009}
  (\bibinfo {year} {2013})}\BibitemShut {NoStop}%
\bibitem [{\citenamefont {Canuel}\ \emph {et~al.}(2006)\citenamefont {Canuel},
  \citenamefont {Leduc}, \citenamefont {Holleville}, \citenamefont {Gauguet},
  \citenamefont {Fils}, \citenamefont {Virdis}, \citenamefont {Clairon},
  \citenamefont {Dimarcq}, \citenamefont {Bord\'e}, \citenamefont {Landragin},\
  and\ \citenamefont {Bouyer}}]{canuel_prl_2006}%
  \BibitemOpen
  \bibfield  {author} {\bibinfo {author} {\bibfnamefont {B.}~\bibnamefont
  {Canuel}}, \bibinfo {author} {\bibfnamefont {F.}~\bibnamefont {Leduc}},
  \bibinfo {author} {\bibfnamefont {D.}~\bibnamefont {Holleville}}, \bibinfo
  {author} {\bibfnamefont {A.}~\bibnamefont {Gauguet}}, \bibinfo {author}
  {\bibfnamefont {J.}~\bibnamefont {Fils}}, \bibinfo {author} {\bibfnamefont
  {A.}~\bibnamefont {Virdis}}, \bibinfo {author} {\bibfnamefont
  {A.}~\bibnamefont {Clairon}}, \bibinfo {author} {\bibfnamefont
  {N.}~\bibnamefont {Dimarcq}}, \bibinfo {author} {\bibfnamefont {C.~J.}\
  \bibnamefont {Bord\'e}}, \bibinfo {author} {\bibfnamefont {A.}~\bibnamefont
  {Landragin}}, \ and\ \bibinfo {author} {\bibfnamefont {P.}~\bibnamefont
  {Bouyer}},\ }\href {\doibase 10.1103/PhysRevLett.97.010402} {\bibfield
  {journal} {\bibinfo  {journal} {Phys. Rev. Lett.}\ }\textbf {\bibinfo
  {volume} {97}},\ \bibinfo {pages} {010402} (\bibinfo {year}
  {2006})}\BibitemShut {NoStop}%
\bibitem [{\citenamefont {Tino}\ and\ \citenamefont
  {Vetrano}(2007)}]{tino_cqg_2007}%
  \BibitemOpen
  \bibfield  {author} {\bibinfo {author} {\bibfnamefont {G.~M.}\ \bibnamefont
  {Tino}}\ and\ \bibinfo {author} {\bibfnamefont {F.}~\bibnamefont {Vetrano}},\
  }\href {\doibase 10.1088/0264-9381/24/9/001} {\bibfield  {journal} {\bibinfo
  {journal} {Class. Quantum Grav.}\ }\textbf {\bibinfo {volume} {24}},\
  \bibinfo {pages} {2167} (\bibinfo {year} {2007})}\BibitemShut {NoStop}%
\bibitem [{\citenamefont {Dimopoulos}\ \emph {et~al.}(2008)\citenamefont
  {Dimopoulos}, \citenamefont {Graham}, \citenamefont {Hogan}, \citenamefont
  {Kasevich},\ and\ \citenamefont {Rajendran}}]{dimopoulos_prd_2008}%
  \BibitemOpen
  \bibfield  {author} {\bibinfo {author} {\bibfnamefont {S.}~\bibnamefont
  {Dimopoulos}}, \bibinfo {author} {\bibfnamefont {P.~W.}\ \bibnamefont
  {Graham}}, \bibinfo {author} {\bibfnamefont {J.~M.}\ \bibnamefont {Hogan}},
  \bibinfo {author} {\bibfnamefont {M.~A.}\ \bibnamefont {Kasevich}}, \ and\
  \bibinfo {author} {\bibfnamefont {S.}~\bibnamefont {Rajendran}},\ }\href
  {\doibase 10.1103/PhysRevD.78.122002} {\bibfield  {journal} {\bibinfo
  {journal} {Phys. Rev. D}\ }\textbf {\bibinfo {volume} {78}},\ \bibinfo
  {pages} {122002} (\bibinfo {year} {2008})}\BibitemShut {NoStop}%
\bibitem [{\citenamefont {Graham}\ \emph {et~al.}(2016)\citenamefont {Graham},
  \citenamefont {Hogan}, \citenamefont {Kasevich},\ and\ \citenamefont
  {Rajendran}}]{graham_arxiv_2016}%
  \BibitemOpen
  \bibfield  {author} {\bibinfo {author} {\bibfnamefont {P.~W.}\ \bibnamefont
  {Graham}}, \bibinfo {author} {\bibfnamefont {J.~M.}\ \bibnamefont {Hogan}},
  \bibinfo {author} {\bibfnamefont {M.~A.}\ \bibnamefont {Kasevich}}, \ and\
  \bibinfo {author} {\bibfnamefont {S.}~\bibnamefont {Rajendran}},\ }\href@noop
  {} {} (\bibinfo {year} {2016}),\ \Eprint
  {http://arxiv.org/abs/arXiv:1606.01860} {arXiv:1606.01860} \BibitemShut
  {NoStop}%
\bibitem [{\citenamefont {Ho}(1998)}]{ho_prl_1998}%
  \BibitemOpen
  \bibfield  {author} {\bibinfo {author} {\bibfnamefont {T.-L.}\ \bibnamefont
  {Ho}},\ }\href {\doibase 10.1103/PhysRevLett.81.742} {\bibfield  {journal}
  {\bibinfo  {journal} {Phys. Rev. Lett.}\ }\textbf {\bibinfo {volume} {81}},\
  \bibinfo {pages} {742} (\bibinfo {year} {1998})}\BibitemShut {NoStop}%
\bibitem [{\citenamefont {Ohmi}\ and\ \citenamefont
  {Machida}(1998)}]{ohmi_jpsp_1998}%
  \BibitemOpen
  \bibfield  {author} {\bibinfo {author} {\bibfnamefont {T.}~\bibnamefont
  {Ohmi}}\ and\ \bibinfo {author} {\bibfnamefont {K.}~\bibnamefont {Machida}},\
  }\href {\doibase 10.1143/JPSJ.67.1822} {\bibfield  {journal} {\bibinfo
  {journal} {J. Phys. Soc. Jpn}\ }\textbf {\bibinfo {volume} {67}},\ \bibinfo
  {pages} {1822} (\bibinfo {year} {1998})}\BibitemShut {NoStop}%
\bibitem [{\citenamefont {Stamper-Kurn}\ and\ \citenamefont
  {Ueda}(2013)}]{stamper_kurn_rmp_2013}%
  \BibitemOpen
  \bibfield  {author} {\bibinfo {author} {\bibfnamefont {D.~M.}\ \bibnamefont
  {Stamper-Kurn}}\ and\ \bibinfo {author} {\bibfnamefont {M.}~\bibnamefont
  {Ueda}},\ }\href {\doibase 10.1103/RevModPhys.85.1191} {\bibfield  {journal}
  {\bibinfo  {journal} {Rev. Mod. Phys.}\ }\textbf {\bibinfo {volume} {85}},\
  \bibinfo {pages} {1191} (\bibinfo {year} {2013})}\BibitemShut {NoStop}%
\bibitem [{\citenamefont {Kawaguchi}\ and\ \citenamefont
  {Ueda}(2012)}]{kawaguchi_jpr_2012}%
  \BibitemOpen
  \bibfield  {author} {\bibinfo {author} {\bibfnamefont {Y.}~\bibnamefont
  {Kawaguchi}}\ and\ \bibinfo {author} {\bibfnamefont {M.}~\bibnamefont
  {Ueda}},\ }\href {\doibase 10.1016/j.physrep.2012.07.005} {\bibfield
  {journal} {\bibinfo  {journal} {Phys. Rep.}\ }\textbf {\bibinfo {volume}
  {520}},\ \bibinfo {pages} {253} (\bibinfo {year} {2012})}\BibitemShut
  {NoStop}%
\bibitem [{\citenamefont {Sun}\ \emph {et~al.}(1996)\citenamefont {Sun},
  \citenamefont {Fejer}, \citenamefont {Gustafson},\ and\ \citenamefont
  {Byer}}]{sun_prl_1996}%
  \BibitemOpen
  \bibfield  {author} {\bibinfo {author} {\bibfnamefont {K.-X.}\ \bibnamefont
  {Sun}}, \bibinfo {author} {\bibfnamefont {M.~M.}\ \bibnamefont {Fejer}},
  \bibinfo {author} {\bibfnamefont {E.}~\bibnamefont {Gustafson}}, \ and\
  \bibinfo {author} {\bibfnamefont {R.~L.}\ \bibnamefont {Byer}},\ }\href
  {\doibase 10.1103/PhysRevLett.76.3053} {\bibfield  {journal} {\bibinfo
  {journal} {Phys. Rev. Lett.}\ }\textbf {\bibinfo {volume} {76}},\ \bibinfo
  {pages} {3053} (\bibinfo {year} {1996})}\BibitemShut {NoStop}%
\bibitem [{\citenamefont {Eberle}\ \emph {et~al.}(2010)\citenamefont {Eberle},
  \citenamefont {Steinlechner}, \citenamefont {Bauchrowitz}, \citenamefont
  {H\"andchen}, \citenamefont {Vahlbruch}, \citenamefont {Mehmet},
  \citenamefont {M\"uller-Ebhardt},\ and\ \citenamefont
  {Schnabel}}]{eberle_prl_2010}%
  \BibitemOpen
  \bibfield  {author} {\bibinfo {author} {\bibfnamefont {T.}~\bibnamefont
  {Eberle}}, \bibinfo {author} {\bibfnamefont {S.}~\bibnamefont
  {Steinlechner}}, \bibinfo {author} {\bibfnamefont {J.}~\bibnamefont
  {Bauchrowitz}}, \bibinfo {author} {\bibfnamefont {V.}~\bibnamefont
  {H\"andchen}}, \bibinfo {author} {\bibfnamefont {H.}~\bibnamefont
  {Vahlbruch}}, \bibinfo {author} {\bibfnamefont {M.}~\bibnamefont {Mehmet}},
  \bibinfo {author} {\bibfnamefont {H.}~\bibnamefont {M\"uller-Ebhardt}}, \
  and\ \bibinfo {author} {\bibfnamefont {R.}~\bibnamefont {Schnabel}},\ }\href
  {\doibase 10.1103/PhysRevLett.104.251102} {\bibfield  {journal} {\bibinfo
  {journal} {Phys. Rev. Lett.}\ }\textbf {\bibinfo {volume} {104}},\ \bibinfo
  {pages} {251102} (\bibinfo {year} {2010})}\BibitemShut {NoStop}%
\bibitem [{\citenamefont {Mizuno}\ \emph {et~al.}(1997)\citenamefont {Mizuno},
  \citenamefont {R\"{u}diger}, \citenamefont {Schilling}, \citenamefont
  {Winkler},\ and\ \citenamefont {Danzmann}}]{mizuno_oc_1997}%
  \BibitemOpen
  \bibfield  {author} {\bibinfo {author} {\bibfnamefont {J.}~\bibnamefont
  {Mizuno}}, \bibinfo {author} {\bibfnamefont {A.}~\bibnamefont {R\"{u}diger}},
  \bibinfo {author} {\bibfnamefont {R.}~\bibnamefont {Schilling}}, \bibinfo
  {author} {\bibfnamefont {W.}~\bibnamefont {Winkler}}, \ and\ \bibinfo
  {author} {\bibfnamefont {K.}~\bibnamefont {Danzmann}},\ }\href {\doibase
  10.1016/s0030-4018(97)00056-4} {\bibfield  {journal} {\bibinfo  {journal}
  {Opt. Commun.}\ }\textbf {\bibinfo {volume} {138}},\ \bibinfo {pages} {383}
  (\bibinfo {year} {1997})}\BibitemShut {NoStop}%
\bibitem [{\citenamefont {Punturo}\ \emph {et~al.}(2010)\citenamefont {Punturo}
  \emph {et~al.}}]{punturoi_cqg_2010}%
  \BibitemOpen
  \bibfield  {author} {\bibinfo {author} {\bibfnamefont {M.}~\bibnamefont
  {Punturo}} \emph {et~al.},\ }\href {\doibase 10.1088/0264-9381/27/8/084007}
  {\bibfield  {journal} {\bibinfo  {journal} {Class. Quantum Grav.}\ }\textbf
  {\bibinfo {volume} {27}},\ \bibinfo {pages} {084007} (\bibinfo {year}
  {2010})}\BibitemShut {NoStop}%
\bibitem [{\citenamefont {Mavalvala}\ \emph {et~al.}(2010)\citenamefont
  {Mavalvala}, \citenamefont {McClelland}, \citenamefont {Mueller},
  \citenamefont {Reitze}, \citenamefont {Schnabel},\ and\ \citenamefont
  {Willke}}]{mavalvala_grg_2010}%
  \BibitemOpen
  \bibfield  {author} {\bibinfo {author} {\bibfnamefont {N.}~\bibnamefont
  {Mavalvala}}, \bibinfo {author} {\bibfnamefont {D.~E.}\ \bibnamefont
  {McClelland}}, \bibinfo {author} {\bibfnamefont {G.}~\bibnamefont {Mueller}},
  \bibinfo {author} {\bibfnamefont {D.~H.}\ \bibnamefont {Reitze}}, \bibinfo
  {author} {\bibfnamefont {R.}~\bibnamefont {Schnabel}}, \ and\ \bibinfo
  {author} {\bibfnamefont {B.}~\bibnamefont {Willke}},\ }\href {\doibase
  10.1007/s10714-010-1023-3} {\bibfield  {journal} {\bibinfo  {journal} {Gen
  Relativ Gravit}\ }\textbf {\bibinfo {volume} {43}},\ \bibinfo {pages} {569}
  (\bibinfo {year} {2010})}\BibitemShut {NoStop}%
\bibitem [{\citenamefont {Abbott}(2016)}]{PhysRevLett.116.061102}%
  \BibitemOpen
  \bibfield  {author} {\bibinfo {author} {\bibfnamefont {B.~P.}\
  \bibnamefont {Abbott}} (\bibinfo {collaboration} {LIGO Scientific
  Collaboration and Virgo Collaboration}),\ }\href {\doibase
  10.1103/PhysRevLett.116.061102} {\bibfield  {journal} {\bibinfo  {journal}
  {Phys. Rev. Lett.}\ }\textbf {\bibinfo {volume} {116}},\ \bibinfo {pages}
  {061102} (\bibinfo {year} {2016})}\BibitemShut {NoStop}%
\bibitem [{\citenamefont {Thanvanthri}\ \emph {et~al.}(2008)\citenamefont
  {Thanvanthri}, \citenamefont {Kapale},\ and\ \citenamefont
  {Dowling}}]{PhysRevA.77.053825}%
   \BibitemOpen
  \bibfield  {author} {\bibinfo {author} {\bibfnamefont {S.}~\bibnamefont
  {Thanvanthri}}, \bibinfo {author} {\bibfnamefont {K.~T.}\ \bibnamefont
  {Kapale}}, \ and\ \bibinfo {author} {\bibfnamefont {J.~P.}\ \bibnamefont
  {Dowling}},\ }\href {\doibase 10.1103/PhysRevA.77.053825} {\bibfield
  {journal} {\bibinfo  {journal} {Phys. Rev. A}\ }\textbf {\bibinfo {volume}
  {77}},\ \bibinfo {pages} {053825} (\bibinfo {year} {2008})}\BibitemShut
  {NoStop}%
\bibitem [{\citenamefont {Kapale}\ and\ \citenamefont
  {Dowling}(2005)}]{PhysRevLett.95.173601}%
  \BibitemOpen
  \bibfield  {author} {\bibinfo {author} {\bibfnamefont {K.~T.}\ \bibnamefont
  {Kapale}}\ and\ \bibinfo {author} {\bibfnamefont {J.~P.}\ \bibnamefont
  {Dowling}},\ }\href {\doibase 10.1103/PhysRevLett.95.173601} {\bibfield
  {journal} {\bibinfo  {journal} {Phys. Rev. Lett.}\ }\textbf {\bibinfo
  {volume} {95}},\ \bibinfo {pages} {173601} (\bibinfo {year}
  {2005})}\BibitemShut {NoStop}%
\bibitem [{\citenamefont {Isoshima}\ \emph {et~al.}(2000)\citenamefont
  {Isoshima}, \citenamefont {Nakahara}, \citenamefont {Ohmi},\ and\
  \citenamefont {Machida}}]{isoshima_pra_2000}%
  \BibitemOpen
  \bibfield  {author} {\bibinfo {author} {\bibfnamefont {T.}~\bibnamefont
  {Isoshima}}, \bibinfo {author} {\bibfnamefont {M.}~\bibnamefont {Nakahara}},
  \bibinfo {author} {\bibfnamefont {T.}~\bibnamefont {Ohmi}}, \ and\ \bibinfo
  {author} {\bibfnamefont {K.}~\bibnamefont {Machida}},\ }\href {\doibase
  10.1103/PhysRevA.61.063610} {\bibfield  {journal} {\bibinfo  {journal} {Phys.
  Rev. A}\ }\textbf {\bibinfo {volume} {61}},\ \bibinfo {pages} {063610}
  (\bibinfo {year} {2000})}\BibitemShut {NoStop}%
\bibitem [{\citenamefont {Ogawa}\ \emph {et~al.}(2002)\citenamefont {Ogawa},
  \citenamefont {M\"ott\"onen}, \citenamefont {Nakahara}, \citenamefont
  {Ohmi},\ and\ \citenamefont {Shimada}}]{ogawa_pra_2002}%
  \BibitemOpen
  \bibfield  {author} {\bibinfo {author} {\bibfnamefont {S.-I.}\ \bibnamefont
  {Ogawa}}, \bibinfo {author} {\bibfnamefont {M.}~\bibnamefont {M\"ott\"onen}},
  \bibinfo {author} {\bibfnamefont {M.}~\bibnamefont {Nakahara}}, \bibinfo
  {author} {\bibfnamefont {T.}~\bibnamefont {Ohmi}}, \ and\ \bibinfo {author}
  {\bibfnamefont {H.}~\bibnamefont {Shimada}},\ }\href {\doibase
  10.1103/PhysRevA.66.013617} {\bibfield  {journal} {\bibinfo  {journal} {Phys.
  Rev. A}\ }\textbf {\bibinfo {volume} {66}},\ \bibinfo {pages} {013617}
  (\bibinfo {year} {2002})}\BibitemShut {NoStop}%
\bibitem [{\citenamefont {Xu}\ \emph {et~al.}(2010)\citenamefont {Xu},
  \citenamefont {Zhang}, \citenamefont {L\"u},\ and\ \citenamefont
  {You}}]{xu_pra_2010}%
  \BibitemOpen
  \bibfield  {author} {\bibinfo {author} {\bibfnamefont {Z.~F.}\ \bibnamefont
  {Xu}}, \bibinfo {author} {\bibfnamefont {P.}~\bibnamefont {Zhang}}, \bibinfo
  {author} {\bibfnamefont {R.}~\bibnamefont {L\"u}}, \ and\ \bibinfo {author}
  {\bibfnamefont {L.}~\bibnamefont {You}},\ }\href {\doibase
  10.1103/PhysRevA.81.053619} {\bibfield  {journal} {\bibinfo  {journal} {Phys.
  Rev. A}\ }\textbf {\bibinfo {volume} {81}},\ \bibinfo {pages} {053619}
  (\bibinfo {year} {2010})}\BibitemShut {NoStop}%
\bibitem [{\citenamefont {Lovegrove}\ \emph {et~al.}(2016)\citenamefont
  {Lovegrove}, \citenamefont {Borgh},\ and\ \citenamefont
  {Ruostekoski}}]{borgh_pra_2016}%
  \BibitemOpen
  \bibfield  {author} {\bibinfo {author} {\bibfnamefont {J.}~\bibnamefont
  {Lovegrove}}, \bibinfo {author} {\bibfnamefont {M.~O.}\ \bibnamefont
  {Borgh}}, \ and\ \bibinfo {author} {\bibfnamefont {J.}~\bibnamefont
  {Ruostekoski}},\ }\href {\doibase 10.1103/PhysRevA.93.033633} {\bibfield
  {journal} {\bibinfo  {journal} {Phys. Rev. A}\ }\textbf {\bibinfo {volume}
  {93}},\ \bibinfo {pages} {033633} (\bibinfo {year} {2016})}\BibitemShut
  {NoStop}%
\bibitem [{\citenamefont {Hamner}\ \emph {et~al.}(2011)\citenamefont {Hamner},
  \citenamefont {Chang}, \citenamefont {Engels},\ and\ \citenamefont
  {Hoefer}}]{PhysRevLett.106.065302}%
  \BibitemOpen
  \bibfield  {author} {\bibinfo {author} {\bibfnamefont {C.}~\bibnamefont
  {Hamner}}, \bibinfo {author} {\bibfnamefont {J.~J.}\ \bibnamefont {Chang}},
  \bibinfo {author} {\bibfnamefont {P.}~\bibnamefont {Engels}}, \ and\ \bibinfo
  {author} {\bibfnamefont {M.~A.}\ \bibnamefont {Hoefer}},\ }\href {\doibase
  10.1103/PhysRevLett.106.065302} {\bibfield  {journal} {\bibinfo  {journal}
  {Phys. Rev. Lett.}\ }\textbf {\bibinfo {volume} {106}},\ \bibinfo {pages}
  {065302} (\bibinfo {year} {2011})}\BibitemShut {NoStop}%
\bibitem [{\citenamefont {Haine}(2016)}]{haine_prl_2016}%
  \BibitemOpen
  \bibfield  {author} {\bibinfo {author} {\bibfnamefont {S.~A.}\ \bibnamefont
  {Haine}},\ }\href {\doibase 10.1103/PhysRevLett.116.230404} {\bibfield
  {journal} {\bibinfo  {journal} {Phys. Rev. Lett.}\ }\textbf {\bibinfo
  {volume} {116}},\ \bibinfo {pages} {230404} (\bibinfo {year}
  {2016})}\BibitemShut {NoStop}%
\bibitem [{\citenamefont {Ragole}\ and\ \citenamefont
  {Taylor}(2016)}]{Ragole_prl_2016}%
  \BibitemOpen
  \bibfield  {author} {\bibinfo {author} {\bibfnamefont {S.}~\bibnamefont
  {Ragole}}\ and\ \bibinfo {author} {\bibfnamefont {J.~M.}\ \bibnamefont
  {Taylor}},\ }\href {\doibase 10.1103/PhysRevLett.117.203002} {\bibfield
  {journal} {\bibinfo  {journal} {Phys. Rev. Lett.}\ }\textbf {\bibinfo
  {volume} {117}},\ \bibinfo {pages} {203002} (\bibinfo {year}
  {2016})}\BibitemShut {NoStop}%
\bibitem [{\citenamefont {Eckel}\ \emph {et~al.}(2014)\citenamefont {Eckel},
  \citenamefont {Jendrzejewski}, \citenamefont {Kumar}, \citenamefont {Lobb},\
  and\ \citenamefont {Campbell}}]{PhysRevX.4.031052}%
  \BibitemOpen
  \bibfield  {author} {\bibinfo {author} {\bibfnamefont {S.}~\bibnamefont
  {Eckel}}, \bibinfo {author} {\bibfnamefont {F.}~\bibnamefont
  {Jendrzejewski}}, \bibinfo {author} {\bibfnamefont {A.}~\bibnamefont
  {Kumar}}, \bibinfo {author} {\bibfnamefont {C.~J.}\ \bibnamefont {Lobb}}, \
  and\ \bibinfo {author} {\bibfnamefont {G.~K.}\ \bibnamefont {Campbell}},\
  }\href {\doibase 10.1103/PhysRevX.4.031052} {\bibfield  {journal} {\bibinfo
  {journal} {Phys. Rev. X}\ }\textbf {\bibinfo {volume} {4}},\ \bibinfo {pages}
  {031052} (\bibinfo {year} {2014})}\BibitemShut {NoStop}%
\bibitem [{\citenamefont {Jendrzejewski}\ \emph {et~al.}(2014)\citenamefont
  {Jendrzejewski}, \citenamefont {Eckel}, \citenamefont {Murray}, \citenamefont
  {Lanier}, \citenamefont {Edwards}, \citenamefont {Lobb},\ and\ \citenamefont
  {Campbell}}]{PhysRevLett.113.045305}%
  \BibitemOpen
  \bibfield  {author} {\bibinfo {author} {\bibfnamefont {F.}~\bibnamefont
  {Jendrzejewski}}, \bibinfo {author} {\bibfnamefont {S.}~\bibnamefont
  {Eckel}}, \bibinfo {author} {\bibfnamefont {N.}~\bibnamefont {Murray}},
  \bibinfo {author} {\bibfnamefont {C.}~\bibnamefont {Lanier}}, \bibinfo
  {author} {\bibfnamefont {M.}~\bibnamefont {Edwards}}, \bibinfo {author}
  {\bibfnamefont {C.~J.}\ \bibnamefont {Lobb}}, \ and\ \bibinfo {author}
  {\bibfnamefont {G.~K.}\ \bibnamefont {Campbell}},\ }\href {\doibase
  10.1103/PhysRevLett.113.045305} {\bibfield  {journal} {\bibinfo  {journal}
  {Phys. Rev. Lett.}\ }\textbf {\bibinfo {volume} {113}},\ \bibinfo {pages}
  {045305} (\bibinfo {year} {2014})}\BibitemShut {NoStop}%
\bibitem [{\citenamefont {Aftalion}\ and\ \citenamefont
  {Du}(2001)}]{aftalion_pra_2001}%
  \BibitemOpen
  \bibfield  {author} {\bibinfo {author} {\bibfnamefont {A.}~\bibnamefont
  {Aftalion}}\ and\ \bibinfo {author} {\bibfnamefont {Q.}~\bibnamefont {Du}},\
  }\href {\doibase 10.1103/PhysRevA.64.063603} {\bibfield  {journal} {\bibinfo
  {journal} {Phys. Rev. A}\ }\textbf {\bibinfo {volume} {64}},\ \bibinfo
  {pages} {063603} (\bibinfo {year} {2001})}\BibitemShut {NoStop}%
\bibitem [{\citenamefont {Mason}\ and\ \citenamefont
  {Aftalion}(2011)}]{mason_pra_2011}%
  \BibitemOpen
  \bibfield  {author} {\bibinfo {author} {\bibfnamefont {P.}~\bibnamefont
  {Mason}}\ and\ \bibinfo {author} {\bibfnamefont {A.}~\bibnamefont
  {Aftalion}},\ }\href {\doibase 10.1103/PhysRevA.84.033611} {\bibfield
  {journal} {\bibinfo  {journal} {Phys. Rev. A}\ }\textbf {\bibinfo {volume}
  {84}},\ \bibinfo {pages} {033611} (\bibinfo {year} {2011})}\BibitemShut
  {NoStop}%
\bibitem [{\citenamefont {Bell}\ \emph {et~al.}(2016)\citenamefont {Bell},
  \citenamefont {Glidden}, \citenamefont {Humbert}, \citenamefont {Bromley},
  \citenamefont {Haine}, \citenamefont {Davis}, \citenamefont {Neely},
  \citenamefont {Baker},\ and\ \citenamefont
  {Rubinsztein-Dunlop}}]{HRD_NJP_2016}%
  \BibitemOpen
  \bibfield  {author} {\bibinfo {author} {\bibfnamefont {T.~A.}\ \bibnamefont
  {Bell}}, \bibinfo {author} {\bibfnamefont {J.~A.~P.}\ \bibnamefont
  {Glidden}}, \bibinfo {author} {\bibfnamefont {L.}~\bibnamefont {Humbert}},
  \bibinfo {author} {\bibfnamefont {M.~W.~J.}\ \bibnamefont {Bromley}},
  \bibinfo {author} {\bibfnamefont {S.~A.}\ \bibnamefont {Haine}}, \bibinfo
  {author} {\bibfnamefont {M.~J.}\ \bibnamefont {Davis}}, \bibinfo {author}
  {\bibfnamefont {T.~W.}\ \bibnamefont {Neely}}, \bibinfo {author}
  {\bibfnamefont {M.~A.}\ \bibnamefont {Baker}}, \ and\ \bibinfo {author}
  {\bibfnamefont {H.}~\bibnamefont {Rubinsztein-Dunlop}},\ }\href
  {http://stacks.iop.org/1367-2630/18/i=3/a=035003} {\bibfield  {journal}
  {\bibinfo  {journal} {New J. Phys.}\ }\textbf {\bibinfo {volume} {18}},\
  \bibinfo {pages} {035003} (\bibinfo {year} {2016})}\BibitemShut {NoStop}%
\bibitem [{\citenamefont {Rakonjac}\ \emph {et~al.}(2016)\citenamefont
  {Rakonjac}, \citenamefont {Marchant}, \citenamefont {Billam}, \citenamefont
  {Helm}, \citenamefont {Yu}, \citenamefont {Gardiner},\ and\ \citenamefont
  {Cornish}}]{rakonjac_pra_2016}%
  \BibitemOpen
  \bibfield  {author} {\bibinfo {author} {\bibfnamefont {A.}~\bibnamefont
  {Rakonjac}}, \bibinfo {author} {\bibfnamefont {A.~L.}\ \bibnamefont
  {Marchant}}, \bibinfo {author} {\bibfnamefont {T.~P.}\ \bibnamefont
  {Billam}}, \bibinfo {author} {\bibfnamefont {J.~L.}\ \bibnamefont {Helm}},
  \bibinfo {author} {\bibfnamefont {M.~M.~H.}\ \bibnamefont {Yu}}, \bibinfo
  {author} {\bibfnamefont {S.~A.}\ \bibnamefont {Gardiner}}, \ and\ \bibinfo
  {author} {\bibfnamefont {S.~L.}\ \bibnamefont {Cornish}},\ }\href {\doibase
  10.1103/PhysRevA.93.013607} {\bibfield  {journal} {\bibinfo  {journal} {Phys.
  Rev. A}\ }\textbf {\bibinfo {volume} {93}},\ \bibinfo {pages} {013607}
  (\bibinfo {year} {2016})}\BibitemShut {NoStop}%
\bibitem [{\citenamefont {Ray}\ \emph {et~al.}(2014)\citenamefont {Ray},
  \citenamefont {Ruokokoski}, \citenamefont {Kandel}, \citenamefont
  {M\"{o}tt\"{o}nen},\ and\ \citenamefont {Hall}}]{Ray2014}%
  \BibitemOpen
  \bibfield  {author} {\bibinfo {author} {\bibfnamefont {M.~W.}\ \bibnamefont
  {Ray}}, \bibinfo {author} {\bibfnamefont {E.}~\bibnamefont {Ruokokoski}},
  \bibinfo {author} {\bibfnamefont {S.}~\bibnamefont {Kandel}}, \bibinfo
  {author} {\bibfnamefont {M.}~\bibnamefont {M\"{o}tt\"{o}nen}}, \ and\
  \bibinfo {author} {\bibfnamefont {D.~S.}\ \bibnamefont {Hall}},\ }\href
  {\doibase 10.1038/nature12954} {\bibfield  {journal} {\bibinfo  {journal}
  {Nature}\ }\textbf {\bibinfo {volume} {505}},\ \bibinfo {pages} {657}
  (\bibinfo {year} {2014})}\BibitemShut {NoStop}%
\bibitem [{Note1()}]{Note1}%
    \BibitemOpen
    \bibinfo {note} {\label {foot:stability}For the stated interaction strengths,
    our system is ferromagnetic ($g_{\protect \mathrm {s}}<0$), and its energy is
    minimised for maximal $|\protect \mathbf {\protect \mathaccentV
    {bar}016{F}}|$. The splitting procedure generates a counterflow state for
    which $|\protect \mathbf {\protect \mathaccentV {bar}016{F}}|=0$, and so
    minimises the Hamiltonian energy of anti-ferromagnetic ($g_{\protect \mathrm
    {s}}>0$) systems~\cite {ho_prl_1998}. As such, the counterflow state
    constitutes a metastable excited state for ferromagnetic systems, but would
    be a stable ground state for antiferromagnetic systems (e.g., $^{23}$Na in
    $F=1$ or $^{87}$Rb in $F=2$). However, since $^{87}$Rb is only weakly
    ferromagnetic, the state's instability has a long associated
    timescale.}\BibitemShut {Stop}%
\bibitem [{\citenamefont {Chang}\ \emph {et~al.}(2004)\citenamefont {Chang},
  \citenamefont {Hamley}, \citenamefont {Barrett}, \citenamefont {Sauer},
  \citenamefont {Fortier}, \citenamefont {Zhang}, \citenamefont {You},\ and\
  \citenamefont {Chapman}}]{chang_prl_2004}%
  \BibitemOpen
  \bibfield  {author} {\bibinfo {author} {\bibfnamefont {M.-S.}\ \bibnamefont
  {Chang}}, \bibinfo {author} {\bibfnamefont {C.~D.}\ \bibnamefont {Hamley}},
  \bibinfo {author} {\bibfnamefont {M.~D.}\ \bibnamefont {Barrett}}, \bibinfo
  {author} {\bibfnamefont {J.~A.}\ \bibnamefont {Sauer}}, \bibinfo {author}
  {\bibfnamefont {K.~M.}\ \bibnamefont {Fortier}}, \bibinfo {author}
  {\bibfnamefont {W.}~\bibnamefont {Zhang}}, \bibinfo {author} {\bibfnamefont
  {L.}~\bibnamefont {You}}, \ and\ \bibinfo {author} {\bibfnamefont {M.~S.}\
  \bibnamefont {Chapman}},\ }\href {\doibase 10.1103/PhysRevLett.92.140403}
  {\bibfield  {journal} {\bibinfo  {journal} {Phys. Rev. Lett.}\ }\textbf
  {\bibinfo {volume} {92}},\ \bibinfo {pages} {140403} (\bibinfo {year}
  {2004})}\BibitemShut {NoStop}%
\bibitem [{\citenamefont {Nickolls}\ \emph {et~al.}(2008)\citenamefont
  {Nickolls}, \citenamefont {Buck}, \citenamefont {Garland},\ and\
  \citenamefont {Skadron}}]{cuda_2008}%
  \BibitemOpen
  \bibfield  {author} {\bibinfo {author} {\bibfnamefont {J.}~\bibnamefont
  {Nickolls}}, \bibinfo {author} {\bibfnamefont {I.}~\bibnamefont {Buck}},
  \bibinfo {author} {\bibfnamefont {M.}~\bibnamefont {Garland}}, \ and\
  \bibinfo {author} {\bibfnamefont {K.}~\bibnamefont {Skadron}},\ }\href
  {\doibase 10.1145/1365490.1365500} {\bibfield  {journal} {\bibinfo  {journal}
  {Queue}\ }\textbf {\bibinfo {volume} {6}},\ \bibinfo {pages} {40} (\bibinfo
  {year} {2008})}\BibitemShut {NoStop}%
\bibitem [{\citenamefont {Hahn}(1950)}]{PhysRev.80.580}%
  \BibitemOpen
  \bibfield  {author} {\bibinfo {author} {\bibfnamefont {E.~L.}\ \bibnamefont
  {Hahn}},\ }\href {\doibase 10.1103/PhysRev.80.580} {\bibfield  {journal}
  {\bibinfo  {journal} {Phys. Rev.}\ }\textbf {\bibinfo {volume} {80}},\
  \bibinfo {pages} {580} (\bibinfo {year} {1950})}\BibitemShut {NoStop}%
\bibitem [{\citenamefont {Gross}(2012)}]{gross_jpb_2012}%
  \BibitemOpen
  \bibfield  {author} {\bibinfo {author} {\bibfnamefont {C.}~\bibnamefont
  {Gross}},\ }\href {\doibase 10.1088/0953-4075/45/10/103001} {\bibfield
  {journal} {\bibinfo  {journal} {J Phys. B}\ }\textbf {\bibinfo {volume}
  {45}},\ \bibinfo {pages} {103001} (\bibinfo {year} {2012})}\BibitemShut
  {NoStop}%
\bibitem [{\citenamefont {Ferris}\ and\ \citenamefont
  {Davis}(2010)}]{ferris_njp_2010}%
  \BibitemOpen
  \bibfield  {author} {\bibinfo {author} {\bibfnamefont {A.~J.}\ \bibnamefont
  {Ferris}}\ and\ \bibinfo {author} {\bibfnamefont {M.~J.}\ \bibnamefont
  {Davis}},\ }\href {\doibase 10.1088/1367-2630/12/5/055024} {\bibfield
  {journal} {\bibinfo  {journal} {New J. Phys.}\ }\textbf {\bibinfo {volume}
  {12}},\ \bibinfo {pages} {055024} (\bibinfo {year} {2010})}\BibitemShut
  {NoStop}%
\bibitem [{dat()}]{data}%
  \BibitemOpen
  \href {\doibase 10.15128/r1m900nt41q} {\ 10.15128/r1m900nt41q}\BibitemShut
  {NoStop}%
\end{thebibliography}
\end{document}